\documentclass[11pt,a4paper]{article}
\usepackage[authoryear]{natbib}
\usepackage{times}
\usepackage{graphics}
\usepackage{ifpdf}
\usepackage{graphicx}
\usepackage{amssymb}
\usepackage{amsmath,latexsym,xspace}
\usepackage{array}
\usepackage{setspace} 
\usepackage{comment}
\usepackage{authblk}
\usepackage[scale={0.73,0.86},centering,includeheadfoot]{geometry}

\def\mm#1{\ensuremath{\boldsymbol{#1}}}

\setkeys{Gin}{width=0.45\textwidth}

\usepackage{subcaption}
\usepackage{verbatim}
\usepackage{comment}

\begin{document}
\begin{titlepage}
\title{Fractional Gaussian noise: Prior specification and model
    comparison}
\author[1]{Sigrunn Holbek S\o rbye}
\author[2]{H\aa vard Rue}

\affil[1]{Department of Mathematics and Statistics, UiT The Arctic University of Norway, 9037 Troms{\o}, Norway. e-mail: sigrunn.sorbye@uit.no}
\affil[2]{Department of Mathematical Sciences, Norwegian University of Science and Technology, 7491 Trondheim, Norway. e-mail: hrue@math.ntnu.no}
\date{\today}
\end{titlepage}
\maketitle
\begin{abstract}
    Fractional Gaussian noise (fGn) is a self-similar stochastic
    process used to model anti-persistent or persistent dependency
    structures in observed time series. Properties of the
    autocovariance function of fGn are characterised by the Hurst
    exponent (H), which in Bayesian contexts typically has been
    assigned a uniform prior on the unit interval. This paper argues
    why a uniform prior is unreasonable and introduces the use of a
    penalised complexity (PC) prior for H. The PC prior is computed
    to penalise divergence from the special case of white noise, and
    is invariant to reparameterisations. An immediate advantage is
    that the exact same prior can be used for the autocorrelation
    coefficient  of a first-order autoregressive process AR(1),
    as this model also reflects a flexible version of white noise.
    Within the general setting of latent Gaussian models, this allows
    us to compare an fGn model component with AR(1) using Bayes
    factors, avoiding confounding effects of prior choices for the
    hyperparameters.  Among others, this is useful in
    climate regression models where inference for underlying linear or
    smooth trends depends heavily on the assumed noise model.
\end{abstract}

{\bf Key words}: Autoregressive process; Bayes factor; long-range dependence;
    PC prior; \texttt{R-INLA}
\maketitle
\section{Introduction}

Many real time series exhibit statistical properties which can be
modelled by self-similar stochastic processes. Such processes have
probability distributions which are invariant to changes of scale in
time (or space), and are particularly useful to model long-range
dependency structures. Long-memory is observed within a wide range of
fields, among others economics, climatology, geophysics and network
engineering \citep{beran:13}. Of specific use is fractional Brownian
motion (fBm) \citep{mandelbrot:68}, which is the only
self-similar continuous-time Gaussian process with stationary
increments. Its discrete-time increment process, referred to as
fractional Gaussian noise (fGn), has autocovariance function
characterised by the self-similarity parameter $H\in(0,1)$. This
parameter is often referred to as the Hurst exponent, known to
quantify the Hurst phenomenon \citep{hurst:51}.

Several methods to estimate $H$ have been proposed in the literature,
among others heuristic approaches like the rescaled range method
\citep{hurst:51}, detrended fluctuation analysis \citep{peng:94} and
the rescaled variance method \citep{girartis:03}. Alternative
approaches include use of wavelets \citep{mccoy:96, abry:98}, and
maximum likelihood and Whittle estimation, see \cite{beran:13} for a
comprehensive overview. Using a Bayesian framework, $H$ has typically
been assigned a uniform prior
\citep{benmehdi:11,makarava:11,makarava:12}. This is computationally
simple and in lack of prior knowledge, a non-informative prior might
seem like a good choice. However, one major drawback is that the
uniform prior is not invariant to reparameterisations, and as seen in
Section~\ref{sec:pc-h} this can give surprising results when the prior
is transformed.

This paper introduces the use of penalised complexity (PC) priors
\citep{simpson:16}, in estimating the Hurst exponent $H$. We also use
a PC prior for the precision of the process. PC priors represent a new
principle-based approach to compute priors for a wide range of
hyperparameters in hierarchical models, where a given model component
can be seen as a flexible version of a simple base model. The simplest
base model in the case of fGn is to have no random effect,
corresponding to infinite precision. For a fixed precision, the fGn
process represents a flexible version of white noise ($H=0.5$). The PC
prior is computed to penalise model component complexity. This is
achieved by assigning a prior to a measure of distance from the
flexible model to the base model, which is then transformed to give a
prior for the hyperparameter of interest. The informativeness of the
prior is adjusted by an intuitive and interpretable used-defined
scaling criterion.

The framework of PC priors has several beneficial properties,
including robustness, invariance to reparameterisations, and it also
provides meaningful priors with a clear interpretation
\citep{simpson:16,riebler:16}. This paper utilises the invariance
property to compare fGn with a first-order autoregressive process,
AR(1), using Bayes factor. Similarly to fGn, the AR(1) process
represents a flexible version of uncorrelated white noise, where the
first-lag autocorrelation coefficient $\phi=0$. This implies that the
PC priors for the hyperparameters $H$ and $\phi$ can be chosen as
transformations of the exact same prior assigned to the distance
measure. This eliminates confounding effects of prior choices in the
calculation of Bayes factor which is very important as the Bayes
factor is known to be sensitive to prior choices \citep{raftery:95}. A
comparison of the AR and fGn models is for example relevant in
analysing climatic time series \citep{lovsletten:16}, identifying
short versus long-range dependency structures in temperature series at
different temporal and/or spatial scales. However, the given ideas
could also be used in comparing other models as long as these
represent flexible versions of the same base model.

The structure of this paper is as follows. Section~\ref{sec:pc-h}
gives a brief review of the principles underlying PC priors and
utilise these to derive the PC prior for $H$. Specifically, the model
complexity of fGn is seen to be non-symmetric around $H=0.5$ which
implies that the prior for $H$ should also be non-symmetric.
Section~\ref{sec:pc-h} also illustrates how Beta priors on $H$ would
shift the mode of the prior for distance away from the natural simple
base model. In section~\ref{sec:fgn-ar1}, we focus on model comparison
of fGn and the first-order AR process using Bayes factor. We perform a
simulation study and also apply the given ideas to check for fGn
structure in global land and sea-surface temperatures, respectively.
Both the estimation and the comparison of models can be computed
within the general framework of latent Gaussian models, using the
methodology of integrated nested Laplace approximations (INLA)
\citep{rueal:09}. This implies that inference for fGn models is made
easily available and that fGn can be combined with other model
components in constructing an additive predictor, for example
including covariates and non-linear or oscillatory random effects.
Concluding remarks are given in section~\ref{sec:conclusions}, while
Appendix~\ref{sec:appendix} describes the implementation of fGn using
the \texttt{R}-interface $\texttt{R-INLA}$, combined with generic
functions.

\section{Penalised complexity priors for the parameters of fractional
    Gaussian noise}\label{sec:pc-h}

Prior selection for hyperparameters is a difficult issue in Bayesian
statistics and often subject to rather ad-hoc choices. A
computationally simple choice is to adopt flat, non-informative
priors. In the case of fGn, the Hurst parameter $H$ has typically been
assigned a uniform prior, argued for in terms of having no knowledge
about the parameter \citep{benmehdi:11, makarava:11, makarava:12}.
Also, it is suggested to use the Jeffrey's prior for the marginal
standard deviation of the model, i.e.
$\pi(\tau^{-1/2})\sim \tau^{1/2}.$ In this section, we derive the
penalised complexity prior \citep{simpson:16} for $H$ and suggest to
use a PC prior also for the precision $\tau$. We also describe
drawbacks in using a uniform, or more generally a Beta prior, for $H$.

\subsection{Penalising divergence from a base model}

Define fractional Gaussian noise as a zero-mean multinormal vector
$\mm{x}'=(x_1,\ldots , x_n)\sim N(\mm{0},\tau^{-1}\mm{\Sigma}$). The
correlation matrix $\mm{\Sigma}$ is Toeplitz with first-row elements
\begin{eqnarray*}
  \gamma(k) &= &\frac{1}{2}(|k+1|^{2H}-2|k|^{2H}+|k-1|^{2H}), \quad k=0,\ldots , n-1,
\end{eqnarray*}
where $H\in(0,1)$ is referred to as the Hurst exponent, while $\tau$
denotes the precision parameter. Asymptotically, the autocorrelation
function of fGn is seen to have a power-law decay,
$$\lim_{k\rightarrow \infty} \gamma(k)\sim H(2H-1)|k|^{2(H-1)}.$$
This implies that fGn reduces to uncorrelated white noise when
$H=0.5$. When $H>0.5$, the process has positive correlation reflecting
a persistent autocorrelation structure. Similarly, the autocorrelation
is negative when $H<0.5$, and the resulting process is then referred
to as being anti-persistent.

The specific structure of the autocorrelation function of fGn implies
that the process can be seen as a flexible version of white noise,
where $H$ represents a flexibility parameter. The framework of
penalised complexity priors \citep{simpson:16} takes advantage of this
inherit nested structure, which is seen in many model components where
hyperparameters can be used to define flexible versions of a simple
base model structure. A key idea is to assign a prior to the
divergence from the flexible model to the base model and transform
this to give a prior for the hyperparameter of interest. Specifically,
the prior is calculated based on four principles, stated in
\cite{simpson:16}. These are as follows:
\begin{enumerate}
\item The PC prior is calculated to support Occam's razor, emphasising
    that model simplicity should be preferred over model complexity.
    Specifically, a model component is seen as a flexible version of a
    simple base model specification. In the fGn case, let
    $f_1(H)=\pi(\mm{x}\mid\mm{\Sigma})$ represent the flexible version
    of a white noise base model $f_0=\pi(\mm{x}\mid \mm{I})$, where
    $\mm{I}$ is the identity matrix. The PC prior for $H$ is then
    designed to give shrinkage to the fixed value $H=0.5$.

\item The second principle in deriving a PC prior implies penalisation
    of model complexity using a measure of distance from the flexible
    version of a model component to its base model. This is achieved
    by making use of the Kullback-Leibler divergence, which in the
    zero-mean Gaussian case simplifies to
    $\mbox{KLD}(f_1(H)\parallel f_0) = -\frac{1}{2}\ln|\mm{\Sigma}|$.
    For the fGn process, the determinant will be of order $n$ and we
    can define a distance measure from $f_1$ to $f_0$ as
    \begin{eqnarray}
      d(H) &=& \sqrt{\frac{1}{n}2\mbox{KLD}(f_1(H)\parallel f_0)} =
               \sqrt{-\frac{1}{n}\ln|\mm{\Sigma}|}=  \left\{ 
               \begin{array}{ll}
                 d_1(H), & 0< H< 0.5 \\ 
                 d_2(H), & 0.5\leq H <1.
               \end{array}
                           \right.
                           \label{eq:dist}
    \end{eqnarray}
    which is invariant to $n$.

\item The third principle used to derive PC priors assumes constant
    rate penalisation. This implies that the relative change in the
    prior is constant, independent of the actual value of the distance
    measure. Consequently, the prior assigned to the distance $d$ will
    be exponential. In the fGn case, the PC prior is calculated
    separately for the two distances $d_1(H)$ and $d_2(H)$,
$$\pi(d_i(H))=\frac{1}{2}\lambda \exp(-\lambda d_i(H)),\quad i=1,2,$$
where $H\in(0,1)$ and $\lambda>0$ denotes the rate parameter. The
prior for $H$ is then obtained by an ordinary change of variable
transformation,
\begin{eqnarray}
  \pi_i(H) &  = &  \pi(d_i(H))\left|\frac{d_i(H)}{dH}\right|,
                  \quad i=1,2,\label{eq:prior-fgn}
\end{eqnarray}
where the derivative is found numerically. In practice, truncation due
to the bounded interval of $H$ is avoided by calculating the prior for
the logit transformation $\rho=\mbox{logit}(H)$, which also has a
variance-stabilising effect.

\item The last principle in deriving PC priors states that the rate
    $\lambda$, controlling shrinkage of the prior, should be governed
    by a user-defined and interpretable scaling criterion. This needs
    to be chosen for each specific model component of interest. In the
    given case, a natural and easily-implemented choice is to define a
    tail probability, like
    \begin{eqnarray}
      P(U<H<1)=\alpha\label{eq:u}
    \end{eqnarray}
    where $\alpha$ is a specified small probability. Alternatively,
    the given probability statement could for example define the
    median of the prior. In any of these cases, the corresponding rate
    parameter is
$$\lambda=\frac{-\ln(2\alpha)}{d(U)}.$$
\end{enumerate}

\subsection{Implications using Beta versus PC priors}
\begin{figure}[h]
    \begin{center}
        \rotatebox{270}{\includegraphics[width=6.0cm]{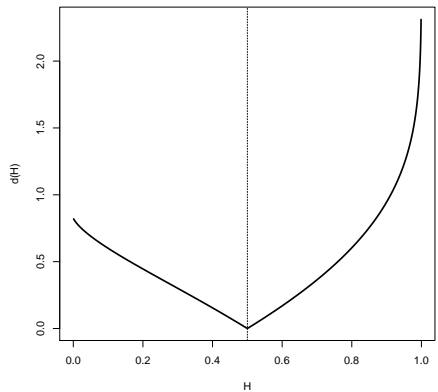}}
        \caption{The distance measure $d(H)$, from a general
            fractional Gaussian process to the white noise base model ($H=0.5$).}
        \label{fig1}
    \end{center}
\end{figure}

The calculated distance function in (\ref{eq:dist}) is non-symmetric
around $H=0.5$ (Figure~\ref{fig1}), clearly illustrating that the
properties of fractional Gaussian noise depend on the value of $H$.
Specifically, the distance measure for volatility or anti-persistent
behaviour ($0<H<0.5$) is seen to decrease more slowly than in the case
of having positive correlation and long memory properties ($0.5<H<1$).
Also, notice that the given distance measure increases rapidly as
$H\rightarrow 0$ and $H\rightarrow 1$, reflecting the major change in
the properties of the fGn process as it approaches its limiting
non-stationary cases.

\begin{figure}[h]
    \begin{center}
        \begin{subfigure}{0.4\textwidth}
            \rotatebox{270}{\includegraphics[width=\linewidth]{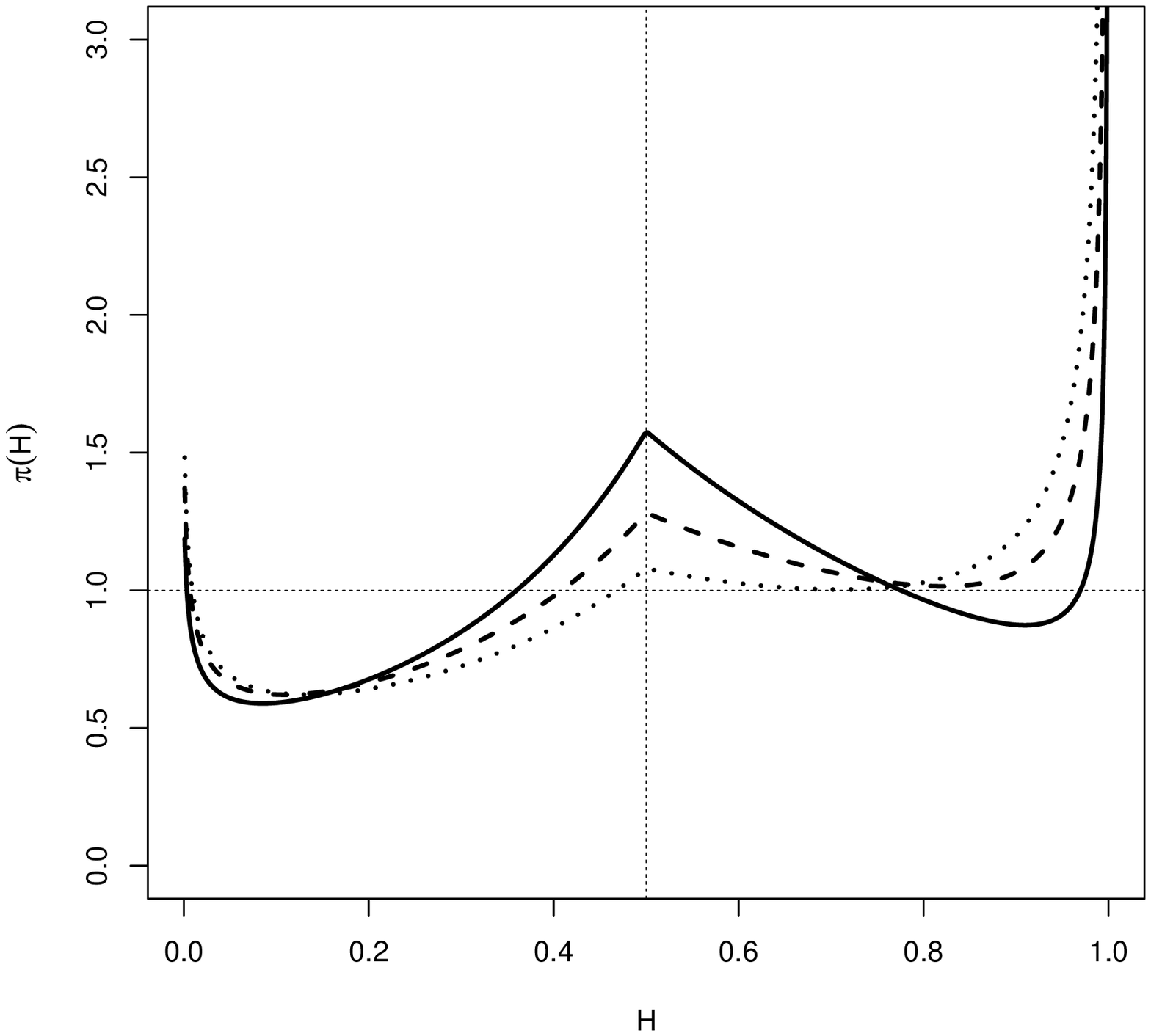}}
            \subcaption{}
        \end{subfigure}
          \makebox[10mm][c]{}
        \begin{subfigure}{0.4\textwidth}
             \rotatebox{270}{\includegraphics[width=\linewidth]{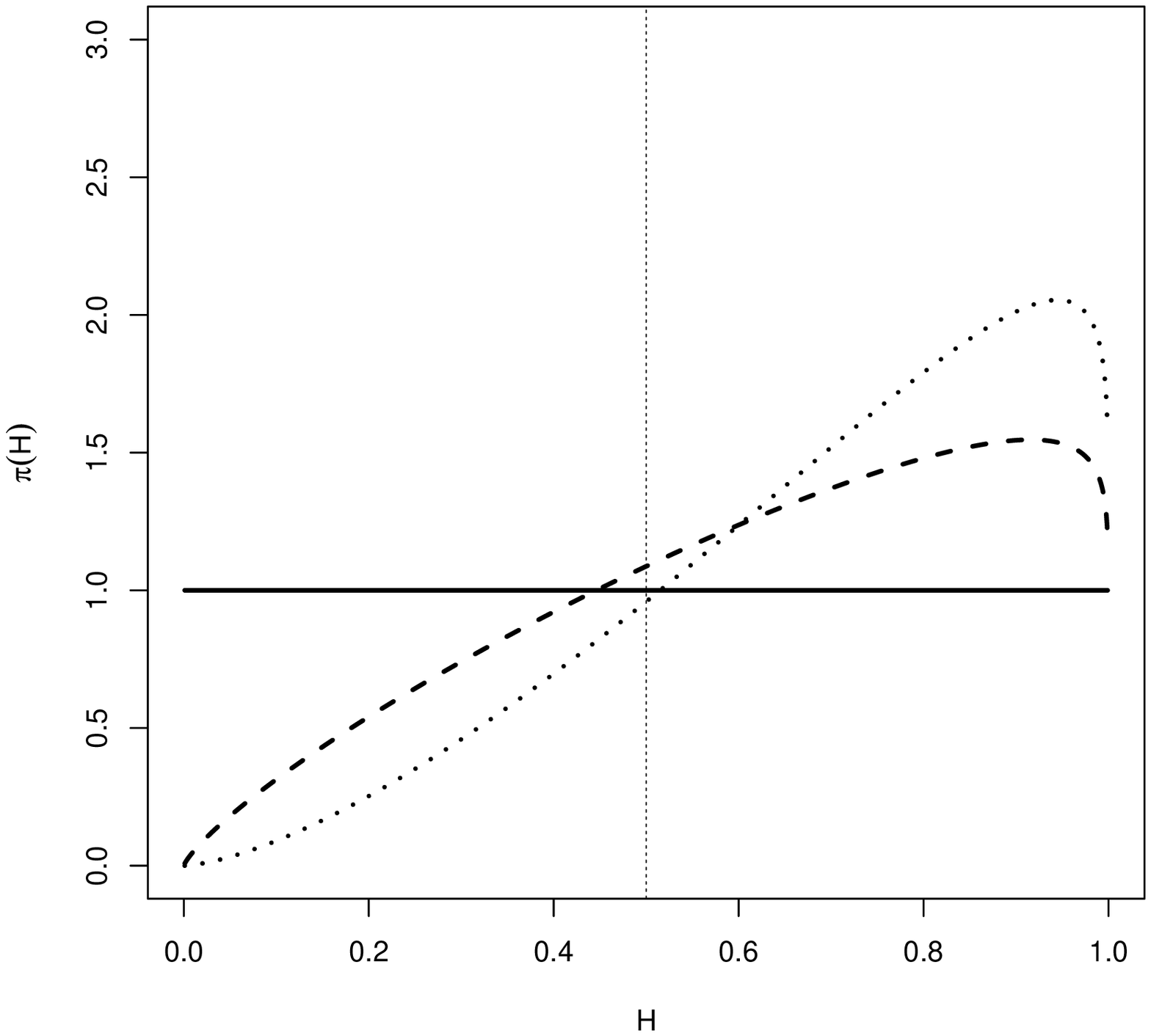}}
            \subcaption{}
        \end{subfigure}
        \begin{subfigure}{0.4\textwidth}
             \rotatebox{270}{\includegraphics[width=\linewidth]{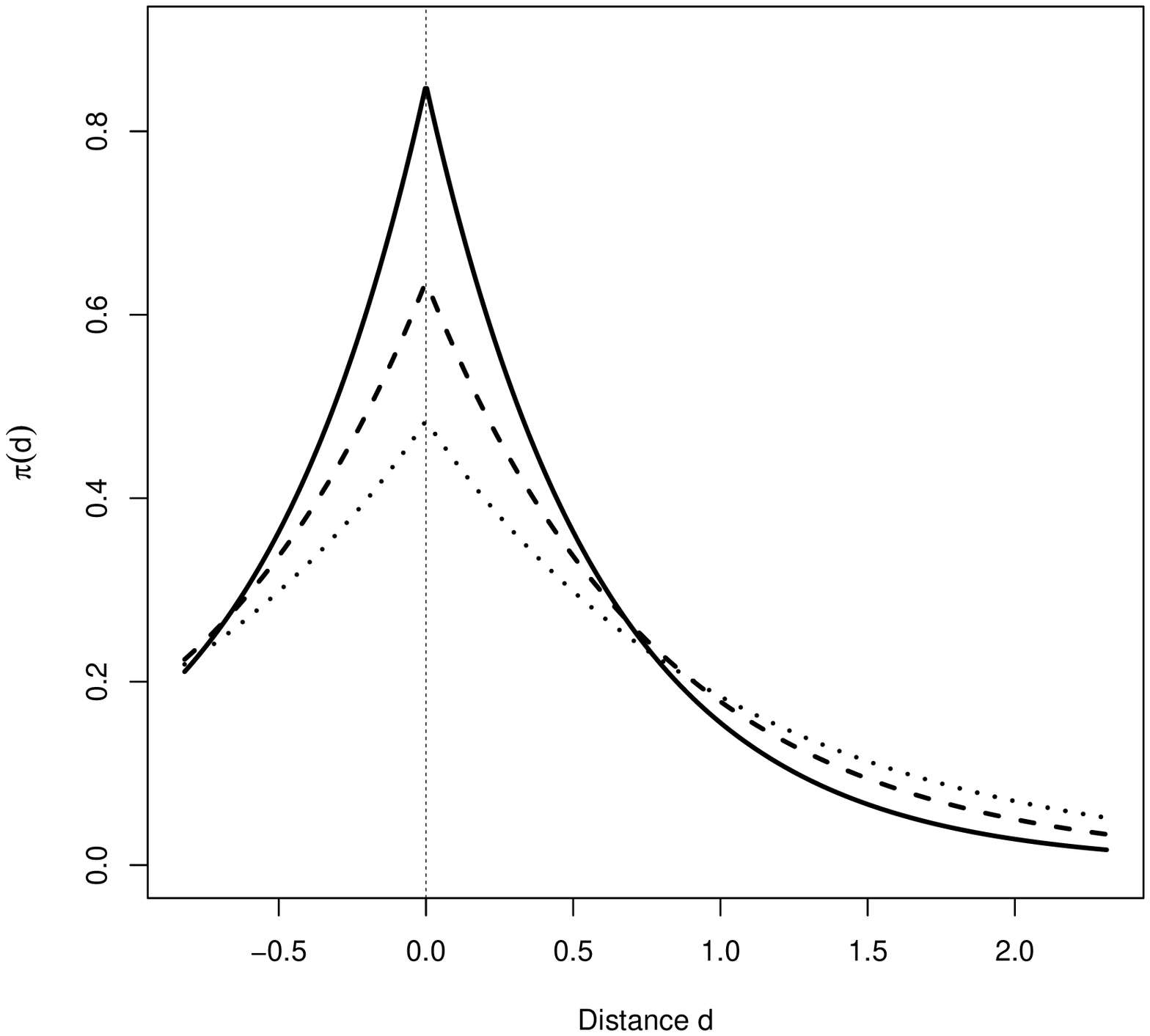}}
            \subcaption{}
        \end{subfigure}
          \makebox[10mm][c]{}
        \begin{subfigure}{0.4\textwidth}
             \rotatebox{270}{\includegraphics[width=\linewidth]{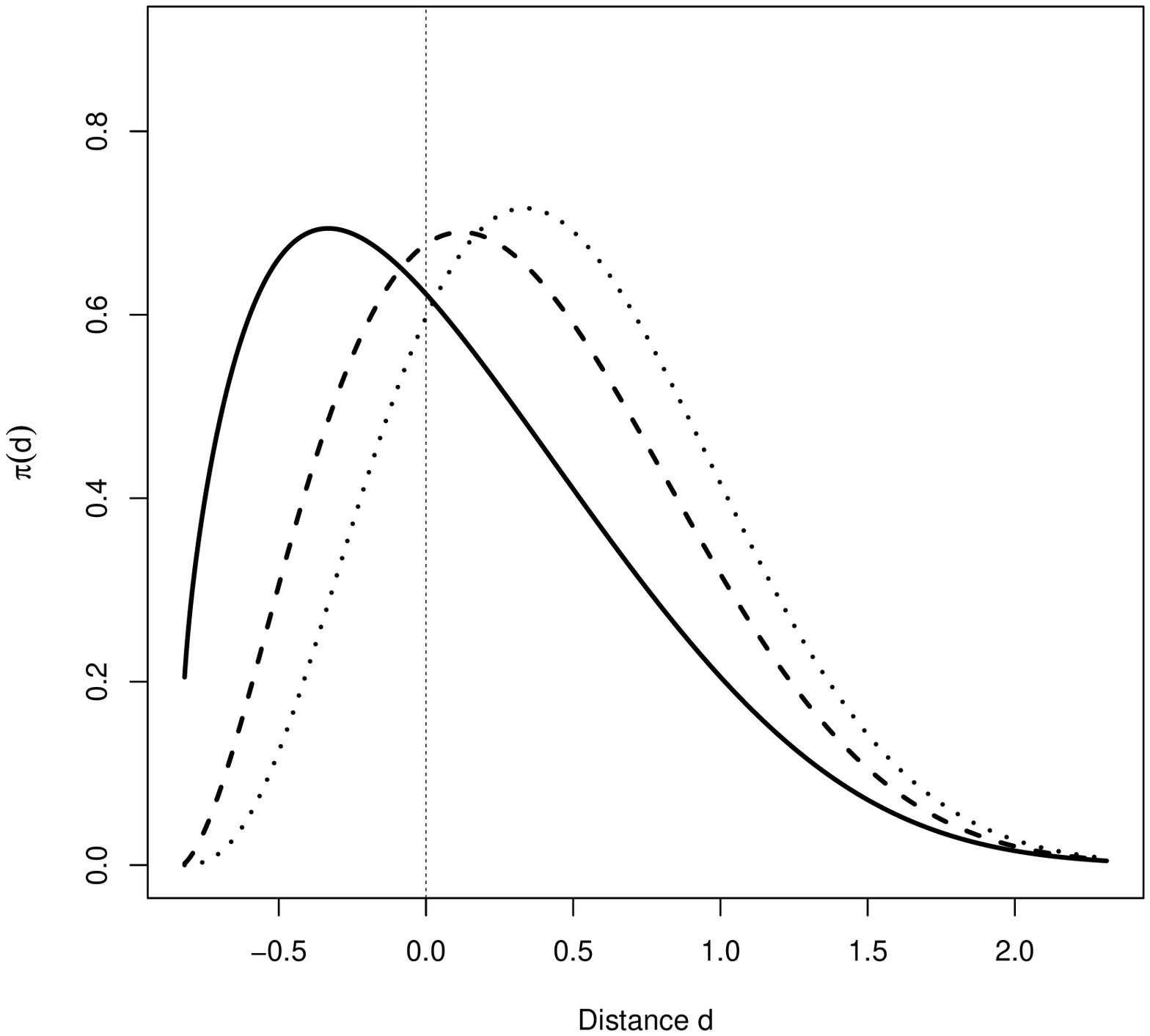}}
            \subcaption{}
        \end{subfigure}
        \caption{Panel a) and b) illustrate PC priors and Beta priors
            for $H$ using three different scalings, in which $P(0.9<H<1)$
            is set equal to 0.1 (solid), 0.15 (dashed) and 0.2
            (dotted), respectively. Panel c) and d) give the corresponding priors
            for the distance measure $d(H)$.} \label{fig2}
    \end{center}
\end{figure}

By transforming the exponential prior on the given distance measure,
the resulting PC prior for $H$ will preserve the properties of the
distance function. Specifically, the PC prior will always have its
mode at the white noise base model ($H=0.5$), independent of the
parameter specifications for $U$ and $\alpha$ in (\ref{eq:u}). Also,
it preserves the non-symmetry of the distance function and it goes to
infinity when $H\rightarrow 0$ and $H\rightarrow 1$. The PC prior for
$H$ is illustrated in Figure~\ref{fig2} (a), using three different
tail probabilities in ~(\ref{eq:u}). Specifically $P(0.9<H<1)$ is set
equal to 0.10, 0.15 and 0.20, in which the last two cases express an
increased prior belief in persistent versus anti-persistent behaviour.
Alternatively, if we knew in advance that a given process has long
memory properties, we could consider to define the prior only for the
interval $H\in (0.5,1)$.

Using the given tail probabilities, the corresponding values for the
rate parameter $\lambda$ of the exponential distribution are
approximately equal to 1.70, 1.27 and 0.97. Figure~\ref{fig2} (b)
illustrates corresponding Beta priors, where the shape parameters
are chosen to give the same tail probability statements as for the PC
priors. Naturally, this can be obtained choosing the shape parameters
of the Beta distribution in numerous way. Here, we just make a
specific choice in which the first shape parameter of the Beta
distribution is set equal to 1.0,1.8 and 2.5, respectively.
Specifically, the first case gives the uniform distribution which
equals the Beta(1,1) case.

In order to understand the implications of different priors on $H$, we
consider the corresponding priors for the distance measure $d(H)$, see
Figure~\ref{fig2} (c) and (d). Here, we use negative distance to
illustrate the priors when $H<0.5$. Notice that when the parameters of
the PC prior are changed, the mode of the prior for distance is still
at the base model, implying that we would not impose a more
complicated model if the true data are in fact white noise. However,
we do change the rate of shrinkage to the base model and increase the
probability of large positive distances. In changing the parameters of
the Beta prior, the mode of the prior for distance is shifted and
pulled away from the base model. Especially, by assigning a uniform
prior to $H$, the corresponding prior on distance is shifted to the
left of the base model. The given negative distance corresponds to a
mode at around $H\approx 0.28$. This seems like an unreasonable
assumption, expect for cases where prior knowledge actually supports
this. For example, theoretical results suggest that the Hurst exponent
is around $1/3$ for various turbulence processes \citep{kang:16}.

\subsection{The PC prior for the precision}

The PC prior for $H$ is derived assuming a fixed precision parameter
$\tau$. An alternative base model for fGn is to have no random effect
($1/\tau=0$). Using the same principles as above, the resulting PC
prior for $\tau$ is the type-2 Gumbel distribution \citep{simpson:16}
\begin{equation}
    \pi(\tau) =\frac{\lambda}{2}\tau^{-3/2}\exp(-\lambda \tau^{-1/2}),
    \quad \lambda>0, \quad \label{eq:prec}
\end{equation}
which prescribes the size of the marginal precision of the fGn
process. This density corresponds to using an exponential prior on the
standard deviation.

Adopting the strategy of \cite{simpson:16}, the rate parameter
$\lambda$ is inferred using the probability statement
$P(1/\sqrt{\tau}>U)=\alpha$, where $\alpha$ is a small probability
while $U$ specifies an upper limit for the marginal standard deviation
$1/\sqrt{\tau}$. For example, if $\alpha=0.01$, the marginal standard
deviation is $0.31U$, after the precision $\tau$ is integrated out. In
practice, the PC prior for the precision is assigned to the
log-precision $\kappa=\log(\tau)$, which is represented in closed form
as
\begin{eqnarray} \pi(\kappa) & =
    &\frac{\lambda}{2}\exp\left(-\lambda\exp(-\frac{\kappa}{2})-
      \frac{\kappa}{2}\right),\quad \quad
    \lambda=-\ln(\alpha)/U.\label{eq:precU}
\end{eqnarray}

\section{Identify long versus short-range dependency using Bayes
    factor}\label{sec:fgn-ar1}

In analysing real time series, it is of major importance to model the
true underlying dependency structure correctly as this will influence
the inference made. Consider a stochastic regression model,
\begin{equation}
    y_t=\nu_t+\epsilon_t,\label{eq:trend}
\end{equation}
where $\nu_t$ represents an underlying smooth trend while the noise
term $\epsilon_t$ models random fluctuations. In analysing temperature
data for different geographical regions, significance of a linear
trend $\nu_t=\beta_0+\beta_1 t$ is seen to depend heavily on whether
the noise term is modelled by fGn or a first-order autoregressive
process \citep{lovsletten:16}.

\subsection{Imposing identical priors for the hyperparameters of fGn
    and first-order AR model}

Autoregressive (AR) processes represent a commonly applied class of
stochastic processes, used to model a one-step or Markovian dependency structure in time
series. Specifically, assume a first-order AR(1) process defined by
$$x_t = \phi x_{t-1}+w_t,\quad t=2,\ldots , n,$$
where the first-lag autocorrelation coefficient $|\phi|<1$. The
innovations are assumed Gaussian, $w_t\sim N(0,\kappa^{-1})$, and
$x_1$ has zero-mean and precision $\tau=\kappa(1-\phi^2)$. The
resulting covariance matrix of $\mm{x}=(x_1,\ldots , x_n)$ is Toeplitz
and defined by the autocorrelation function $\gamma(k)=\phi^k$ for
lags $k=0,\ldots , n-1$.

Similarly to fGn, the first-order AR process represents a flexible
version of uncorrelated white noise ($\phi=0$). The Kullback-Leibler
divergence is of order $n-1$ and an invariant distance measure from
AR(1) to white noise is
$$d(\phi) =\sqrt{-\ln(1-\phi^2)}.$$
The resulting PC prior for $\phi$ \citep{simpson:16} is then expressed
as
\begin{equation}
    \pi(\phi) = \frac{\lambda}{2} \exp
    \left(-\lambda \sqrt{-\ln(1-\phi^{2})}\right) 
    \; \frac{|\phi|}{(1-\phi^2) \sqrt{-\ln(1-\phi^2)}},
    \qquad |\phi|<1\label{eq:prior-ar1}
\end{equation}
where $\lambda>0$.

In general, let $M_0$ and $M_1$ denote two different model components
for the noise term $\mm{\epsilon}$ in (\ref{eq:trend}). Within a
Bayesian framework, the two models can be compared using Bayes factor
which quantifies the evidence in favour of one statistical model. The
Bayes factor is defined as as the ratio of the marginal likelihoods
\begin{eqnarray}
  \frac{\pi(\mm{\epsilon}\mid M_0)}{\pi(\mm{\epsilon}\mid M_1)}
  & = & \frac{\pi(M_0\mid {\mm{\epsilon}})}{\pi(M_1\mid
        {\mm{\epsilon}})}
        \frac{\pi(M_1)}{\pi(M_0)}, \label{eq:bayes-factor}
\end{eqnarray}
where $\pi(M_0)$ and $\pi(M_1)$ represent priors for the two model
components. In general, Bayes factors are sensitive to the choices of
$\pi(M_0)$ and $\pi(M_1)$ \citep{raftery:95}. When the aim is to
compare a model with an fGn component to a model with an AR(1)
component, the Bayes factor will depend on the prior distributions for
the precision parameters of the models and the priors for the
parameters $\phi$ and $H$. The precision parameters have the same
interpretation for both fGn and AR(1), hence we will use the same
prior distribution for the precision of both models. Since we are
using PC priors for both $\phi$ and $H$, these parameters are just a
reparameterisation of the distance from the (unit variance) AR(1) and
fGn model, to the \emph{same} base model. By choosing the same rate
parameter $\lambda$ for the PC priors of both $\phi$ and $H$, the prior
distributions will be the same even if the models and hyperparameters
are different. This is a very convenient feature of PC priors which we
make use of in the following examples.

\subsection{Simulation results}\label{sec:sim}
To illustrate the capability of Bayes factor to identify fGn versus
AR(1) structure, we perform a simulation study and generate fGn
processes of different lengths $n$, for specified values of the Hurst
parameter, scaled to have variance 1. The selected values for the
Hurst parameter are $H=0.7$, $H=0.8$ and $H=0.9$. These all give
persistent fGn processes and are selected to mimic realistic values of
the Hurst parameter in analysing long-memory processes. The PC prior
for $H$ is implemented using the scaling criterion $P(0.9<H<1)=0.10$.
This implies that the rate parameter of the exponential distribution
of the distance measure is $\lambda\approx1.70$. This rate parameter
is used to find the corresponding prior for the first-lag correlation
coefficient in fitting an AR(1) model to the data. The marginal
precisions of both the fGn and AR(1) models are assigned the PC prior
using $U=1$ and $\alpha=0.01$ in~(\ref{eq:precU}).

\begin{table}[tbh]
    \begin{center}
        \begin{tabular}{|ll|ccccc|c|}\hline
          &  & \multicolumn{5}{|c|}{Bayes factor: Strength of evidence} & \\
          & & False & No conclusion & Positive & Strong & Very Strong  & Total\\
          $n$ &  $H$ &$\mbox{BF}<1/3 $ & $1/3<\mbox{BF}<3 $ & $3<\mbox{BF}<20$ &    $20<\mbox{BF}<150$ & $\mbox{BF}>150$ & $\mbox{BF}>3$ \\\hline
          100 & 0.7 & 0.226& 0.685 & 0.077 & 0.009 &0.003 &0.089\\
          200 & 0.7 &  0.232&0.493 &0192 & 0.063& 0.020&0.275 \\
          500 &0.7 &0.097 &0.224 &0.220&0.186 &0.273 &0.679 \\\hline 
          100 & 0.8 & 0.366  & 0.513 & 0.105 & 0.015&  0.001  &  0.121 \\
          200 & 0.8 & 0.260 & 0.353 & 0.219 & 0.123 & 0.045 & 0.387\\
          500 &0.8 & 0.072 & 0.099 &0.121 &0.153 & 0.555 &  0.829\\\hline 
          100 & 0.9 & 0.521 & 0.352 & 0.096 & 0.029 & 0.002 &  0.127 \\
          200 & 0.9 & 0.316 & 0.252 & 0.189 & 0.146 & 0.097 & 0.432\\
          500 &0.9 & 0.060& 0.073& 0.091& 0.104& 0.672&0.867 \\\hline 
        \end{tabular}
    \end{center}
    \label{tab:bf}
    \caption{Estimated proportion of Bayes factors in each of five
        groups, showing strength of evidence of the underlying process
        being fGn.
        Results are based on 1000 simulations for each value of the
        Hurst parameter $H$ and for each time series length $n$.}
\end{table}

In evaluating the evidence given by Bayes factors, we have chosen to
apply the categories given in \cite{raftery:95}, in which Bayes factor
should be larger than 3 to give a positive evidence in favour of one
model. Note that these categories could also be specified in terms of
twice the natural logarithm of the Bayes factor, which would have the
same scale as the likelihood ratio test statistic. Table 1 illustrates
that rather long time series are needed in order for the Bayes factor
to give correct identification of the underlying fGn process. This
seems natural as the long-memory properties might not be clearly
apparent if the time series is short. We also notice that the results
improve with higher values of the Hurst parameter, in which the long-memory
 structure of fGn would get more apparent.
Especially, if $H=0.9$ and the length of the time series is $n=500$,
we have very strong evidence of fGn in two-thirds of the cases.
Naturally, the given results could be improved upon using a prior for
$H$ that puts more probability mass to the upper tail of the
distribution.

\subsection{Temperature data example}\label{sec:real-data}

In this section, we analyse a real temperature data set to investigate
whether we find evidence of fGn compared with AR(1) structure, using
Bayes factor. The dataset "NOAA-CIRES 20th Century Reanalysis V2c", downloaded
from \verb|http://www.esrl.noaa.gov/psd/data/20thC_Rean|, 
contains reanalysed data from the period 1851 - 2014. The data are
available for a 2 degree latitude times a 2 degree longitude global
grid covering the earth  and combine
temperature measurements with model estimated data. 
 
We assume a regression model for temperature as given in
(\ref{eq:trend}), where the trend is assumed to be either linear or
non-linear. The non-linear trend is modelled using an intrinsic CAR
model \citep{besag:91}, also referred to as a second-order intrinsic
Gaussian Markov random field (IGMRF)  defined on a line, see
\citet[Ch.~3]{book80}. The precision parameter of the model is
assigned the penalised complexity prior in (\ref{eq:prec}), using
scaling parameters $(U, \alpha) = (0.10, 0.01)$ in (\ref{eq:precU}). By
choosing a small value of $U$, the prior imposes only small deviation
from a straight line. Note that the given IGMRF model needs to be
scaled to have a generalised variance equal to 1, such that we can use
the same prior for $\tau$ independently of the time scale used for the
analysis \citep{sorbye:14}. The hyperprior specifications for
the AR(1) and fGn components used to model climate noise, are chosen
to be the same as in the simulation study of Section~\ref{sec:sim}.
\begin{table}[tbh]
    \begin{center}
        \begin{tabular}{|l|cc|cc|cc|l|}\hline
            & \multicolumn{2}{|c|}{Trend parameter estimates}  & \multicolumn{4}{|c|}{fGn parameter estimates} &Evidence fGn\\
          Time series & $\hat{\beta}_1$& $95\%$ CI & $\hat H$ & $95\%$ CI & $\hat {\sigma}$ & $95\%$ CI& Category \\\hline
          Land (a)& 0.006 & (0.004, 0.008) & 0.76 & (0.67, 0.84) & 0.22 & (0.19, 0.26) &  Strong\\
          Land (q)   & 0.006 & (0.004, 0.008) & 0.77 & (0.72, 0.82)&  0.31 & (0.29, 0.34)  &  Very strong \\\hline
          Sea (a) & 0.004 & (0.003, 0.006) & 0.95 & (0.91, 0.99) & 0.21 & (0.13, 0.37) & Strong  \\
          Sea (q)  & 0.004 & (0.003, 0.006)  & 0.99& (0.97,1.00) & 0.39 & (0.21, 0.74) & Very strong\\\hline
        \end{tabular}
    \end{center}
    \caption{The mean estimates, $95\%$ credible intervals and the evidence of fGn versus AR(1) given by Bayes factor, fitting a regression model with linear trend to global land and sea temperatures. We use both annual (a) and quarterly (q)  time scales for the years 1851 - 2014.}
    \label{tab:bf-lin}
\end{table}

\begin{table}[tbh]
    \begin{center}
        \begin{tabular}{|l|cc|cc|cc|l|}\hline
            & \multicolumn{2}{|c|}{Trend parameter estimates}  & \multicolumn{4}{|c|}{fGn parameter estimates} &Evidence fGn\\
          Time series & $\hat{\sigma}$& $95\%$ CI & $\hat H$ & $95\%$ CI & $\hat {\sigma}$ & $95\%$ CI& Category \\\hline
          Land (a) & 0.047 & (0.015, 0.102) & 0.69 & (0.58, 0.79) & 0.20 & (0.18, 0.23) & No conclusion \\
          Land (q)  & 0.038 & (0.008, 0.093) & 0.75 & (0.69, 0.80)  & 0.30 & (0.28, 0.33)& Positive  \\\hline
          Sea (a) & 0.041 & (0.012, 0.095) & 0.92 & (0.83, 0.98) & 0.15 & (0.10, 0.23) & No conclusion \\
          Sea (q) & 0.039& (0.012, 0.091) & 0.98 & (0.95, 1.00) & 0.34 & (0.16, 0.66)& No conclusion  \\\hline
        \end{tabular}
    \end{center}
    \caption{The mean estimates, $95\%$ credible intervals and the evidence of fGn versus AR(1) given by Bayes factor, fitting a regression model with a smooth non-linear trend to global land and sea temperatures. We use both annual (a) and quarterly (q)  time scales for the years 1851 - 2014.}
\label{tab:bf-nonlin}
\end{table}

\begin{figure}[h]
    \begin{center}
        \begin{subfigure}{0.4\textwidth}
           \rotatebox{270}{\includegraphics[width=\linewidth]{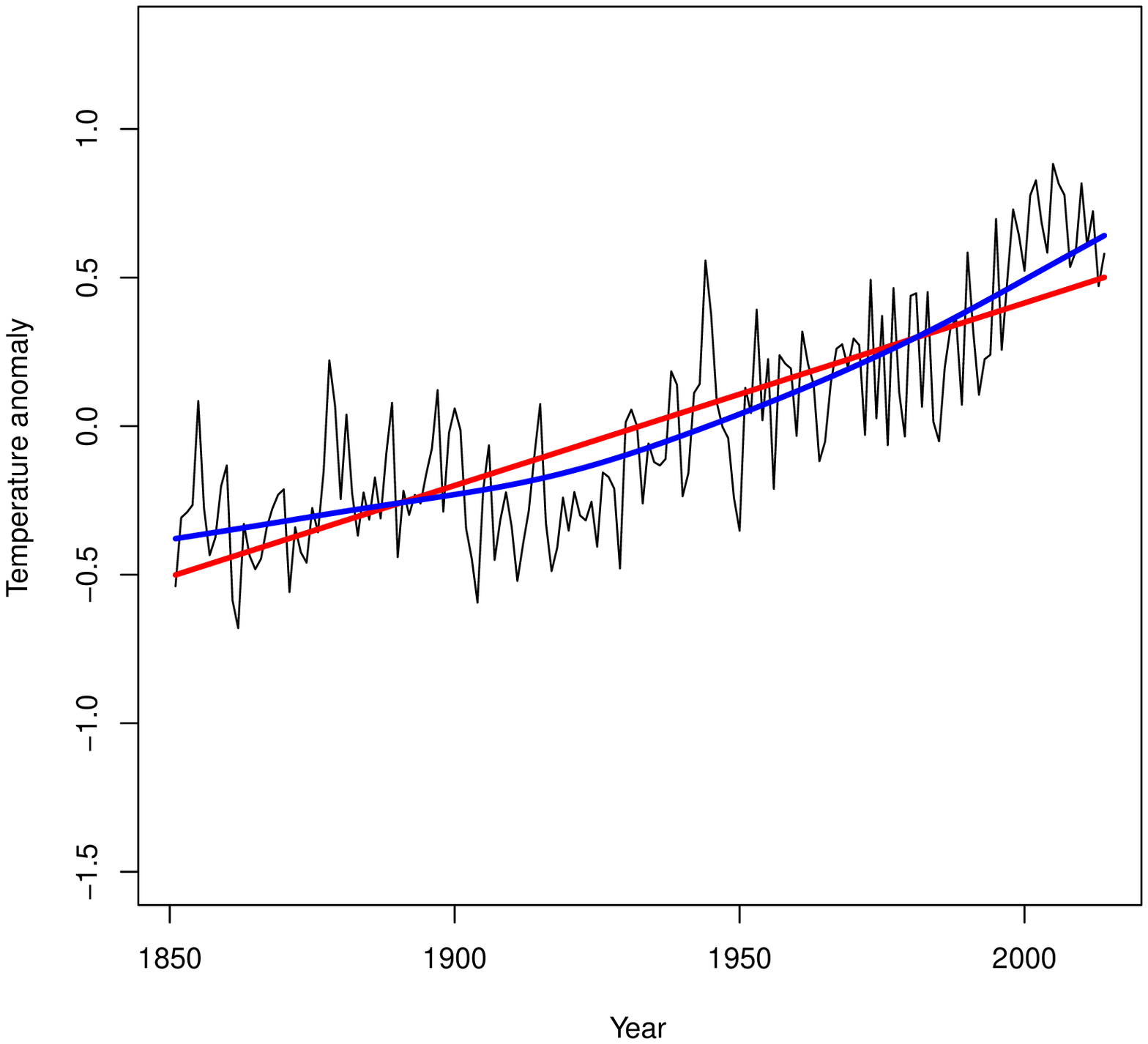}}
            \caption{}
        \end{subfigure}
         \makebox[10mm][c]{}
        \begin{subfigure}{0.4\textwidth}
            \rotatebox{270}{\includegraphics[width=\linewidth]{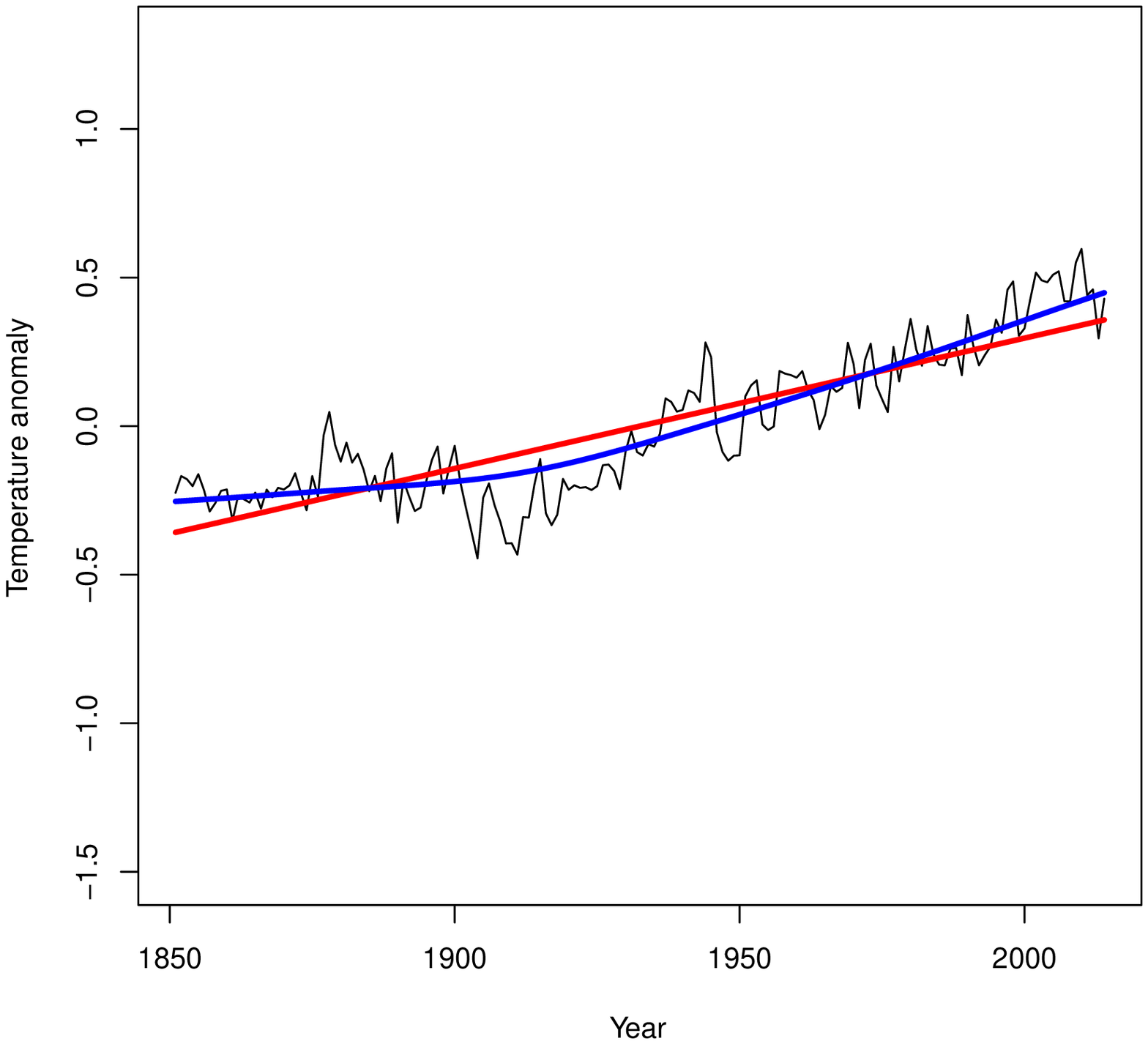}}
            \caption{}
        \end{subfigure}
        \begin{subfigure}{0.4\textwidth}
             \rotatebox{270}{\includegraphics[width=\linewidth]{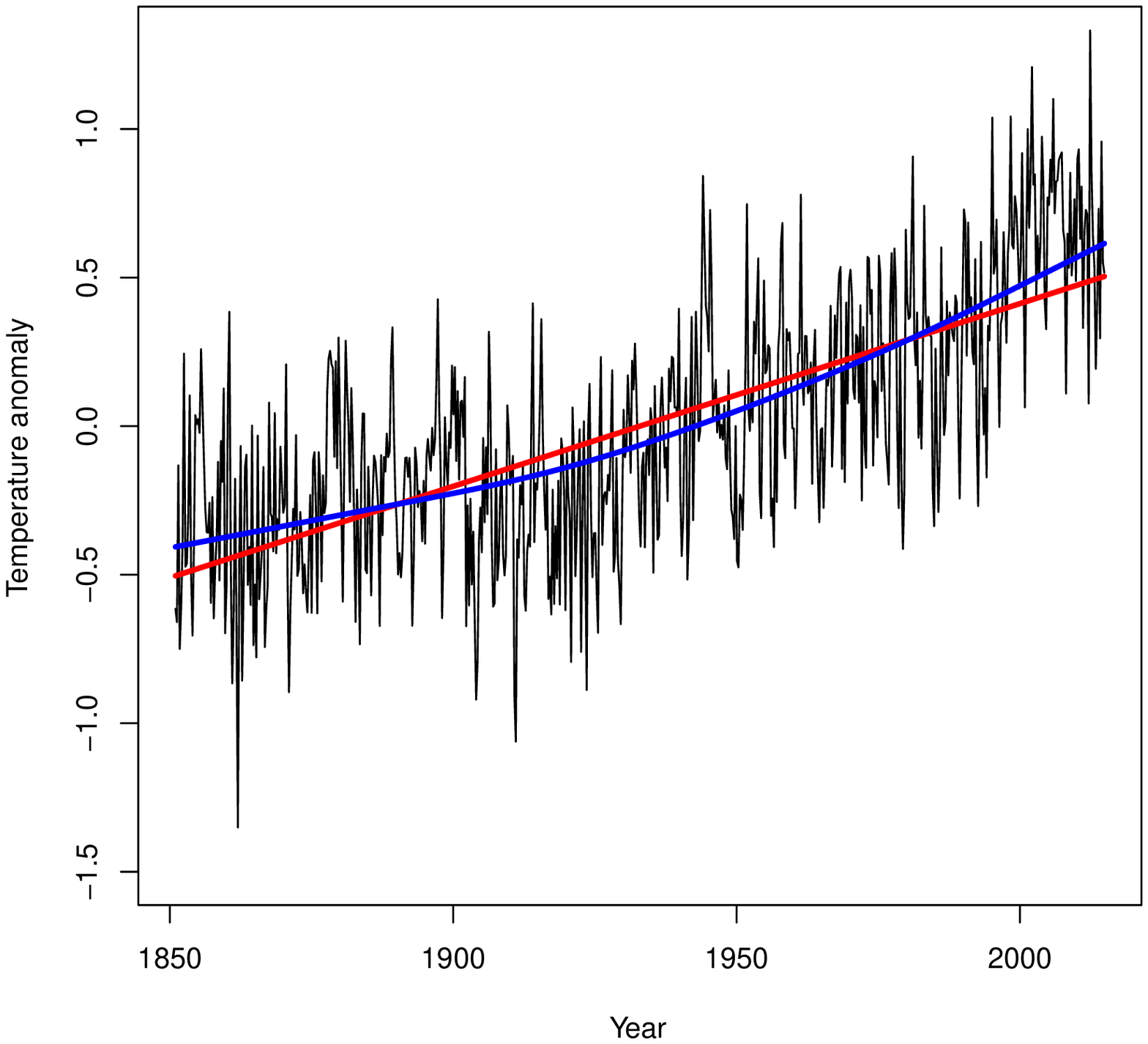}}
            \caption{}
        \end{subfigure}
          \makebox[10mm][c]{}
        \begin{subfigure}{0.4\textwidth}
             \rotatebox{270}{\includegraphics[width=\linewidth]{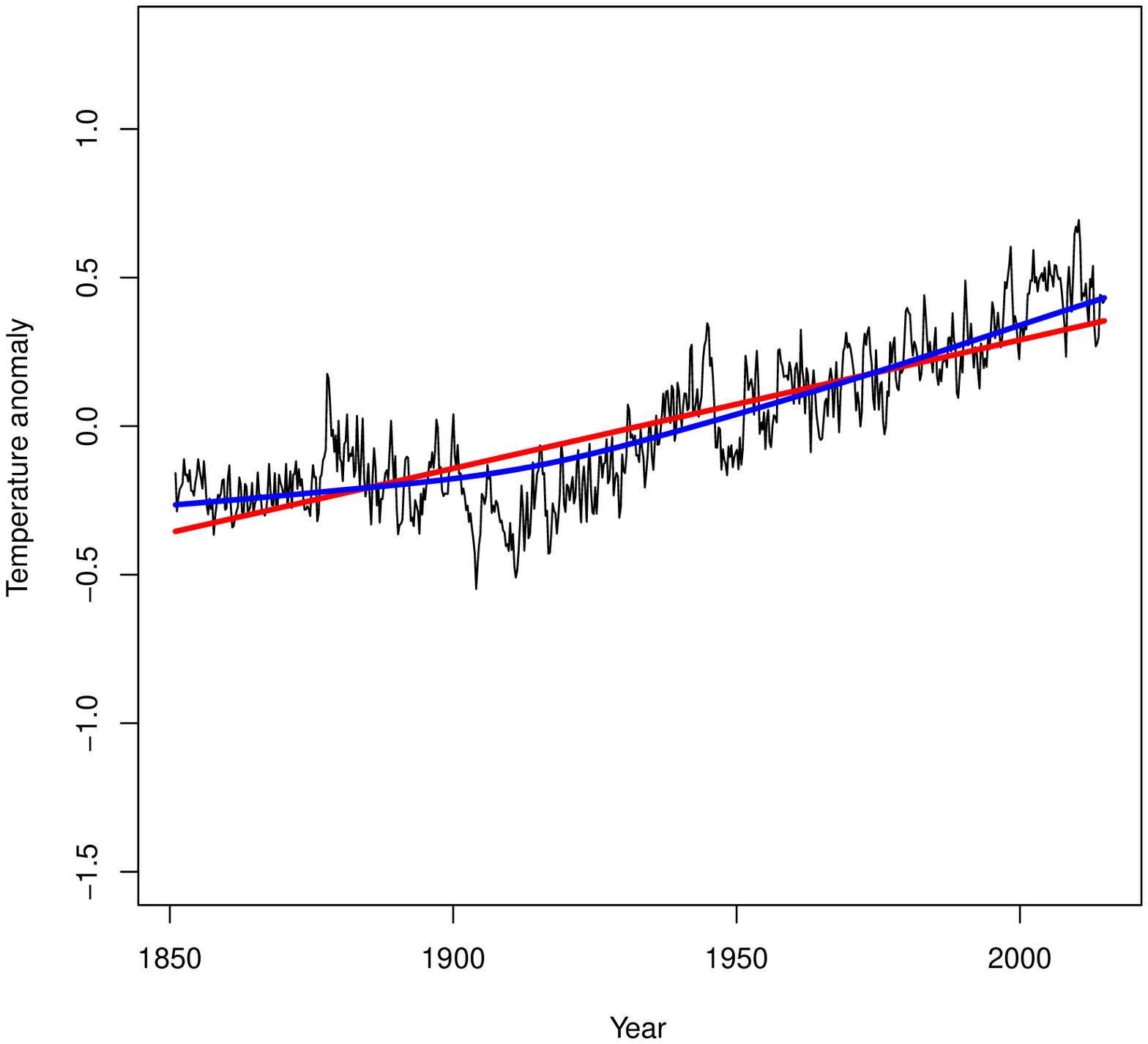}}
            \caption{}
        \end{subfigure}
        \caption{The upper panels show annual global land temperature
            (a) and annual global sea-surface temperature (b) in the period
            1851 - 2014, with estimated linear (red) and non-linear
            (blue) trends. The lower panels show the corresponding
            quarterly global land temperature (c) and sea-surface temperature
            (d).}
        \label{fig-trend}
    \end{center}
\end{figure}

Using the given data, it would be interesting to study evidence of fGn
for different geographical regions, but here we only report results
for aggregated data over land areas and sea-surfaces. We
calculate the Bayes factor (\ref{eq:bayes-factor}) in each case, using
both an annual and quarterly time scale. The results are displayed in
Table~\ref{tab:bf-lin}-\ref{tab:bf-nonlin} using a linear and non-linear trend, respectively. 
The results clearly illustrate that the evidence of fGn
is very sensitive to the trend model. Using a linear trend, the
evidence of fGn compared with AR(1) is strong ($\mbox{BF}>20$) or very strong
($\mbox{BF}>150$). Using a non-linear trend, the test using Bayes factor is
inconclusive, except for the quarterly land data where there is weak
evidence of fGn structure. It is well-known that non-linear trends can
easily be interpreted incorrectly as long memory \citep{beran:13}. The
given results illustrate that a non-linear trend, even a very weak
one (Figure~\ref{fig-trend}), typically captures some of the
long-range dependency structure of the time series, influencing the
Bayes factor. However, it does not change the estimate of $H$,
radically. We also notice that the quarterly data give higher values
for the Hurst parameter and stronger evidence of fGn, than the annual
data, which among others could be explained by the length of these
series being four times longer than for the annual data.

The estimated Hurst parameter is seen to be higher for the sea-surface
temperatures than the land temperatures, both using different trend
models and different time scales. In fact, in the cases where the
Hurst parameter is very close to 1 this could indicate that more
components should be included in the model, for example to explain
oscillatory behaviour. It is well-known that estimation of the Hurst
parameter is not robust against departures from stationarity and
trends and maybe also other models for the climate noise should be
considered. Alternatively, the variability of the trend model could be
adjusted to explain more of the fluctuations of the temperatures. A
big advantage of using PC priors is that such adjustments are easily
made, simply choosing a larger upper value of $U$ in (\ref{eq:precU}).

\section{Concluding remarks}\label{sec:conclusions}
The framework of penalised complexity priors \citep{simpson:16} 
represents a new approach to calculate priors for hyperparameters in Bayesian
hierarchical models. The PC priors are intuitive in the sense that
they are designed to shrink towards a well-defined simple base model and
will always have a local mode at this base model where the distance
measure is 0. Also, the parameters of a PC prior have a clear
interpretation as they govern the rate of shrinkage to the base model
and thereby the informativeness of the prior. In estimating the Hurst
parameter of fractional Gaussian noise, the argument of having no
knowledge about the parameter has encouraged the use of a
non-informative prior. If we accept that a given prior on $H$ should
give reasonable results also on the distance scale, the uniform prior
needs be ruled out. Similarly, changes in the parameters of Beta
priors would shift the mode at distance scale and pull estimates away
from the base model, making interpretation of these parameters
non-intuitive and non-transparent.

In calculating Bayes factor, we take advantage of the fact that PC
priors are invariant to reparameterisations. This implies that priors
for two different hyperparameters simply represent two different
transformations of the same prior on distance scale, as long as the
corresponding model components represent flexible versions of the same
base model. This is very useful in calculation of Bayes factors as
these are known to be rather sensitive to prior choices. In the
current analysis, we have focused on identifying the underlying noise
term in a regression model as either AR(1) or fGn but the given ideas
could also be applied for other model components as long as these
share the same base model. The given analysis demonstrates that it is
difficult to separate the trend from the model noise and that
potential non-linearity in the trend could easily be interpreted as
long memory. This makes it important to control the smoothness of the
underlying trend, in which a linear model is considered as the base
model and deviation from the base model is controlled by a PC prior on
the precision parameter. In general, the use of PC priors is very
helpful to control the effect of different random effects and make the
model components identifiable.

The option to combine generic functions with \texttt{R-INLA}, is a
flexible and easy way to incorporate an fGn model component within the
general class of latent Gaussian models. This makes it possible to
analyse also more complex additive regression models than the ones
presented here, in which an fGn component can be combined with for
example smooth effects of covariates, seasonal and oscillatory
effects, non-linear trend models and also spatial model components.
However, to make full use of the computational power of
\texttt{R-INLA}, the fGn model should be approximated to have Markov
properties as this would give a sparse precision matrix of the
underlying latent field. One alternative to achieve this is to
approximate fGn as a mix of a large number of AR(1) processes,
originally suggested in \cite{granger:80}. Preliminary results
\citep{myrvoll-nilsen:16} indicate that such an approximation is
promising, also using only a few AR(1) components. In fact, we have
already seen that aggregation of just three AR(1) processes give an
excellent approximation of fGn, as long as the weights and
coefficients of the AR components are chosen in a clever way. In
addition, such a decomposition of fGn can potentially be linked to
linear multibox energy balance models \citep{fredriksen:16}, in
analysing temperature data. This requires further study and will be
reported elsewhere.

\section*{Acknowledgement}
The 20th Century Reanalysis V2c data is provided by the NOAA/OAR/ESRL
PSD, Boulder, Colorado, USA, from their Web site at
\texttt{http://www.esrl.noaa.gov/psd/}. The authors wish to thank
Hege-Beate Fredriksen for valuable discussions and for processing the
data to give aggregated land and sea-surface temperatures.

\bibliographystyle{apa} \bibliography{shs,mybib}
\section{Appendix: Incorporating fGn models within latent Gaussian
    models}\label{sec:appendix}

We have implemented the fGn model within the R-INLA package (see
\texttt{www.r-inla.org}) which uses nested Laplace approximations to
do Bayesian inference for the class of latent Gaussian models
\citep{rueal:09,art522}. We used the generic model
``\texttt{rgeneric}'' allowing a (Gaussian) latent model component to
be defined using \texttt{R} and it is passed into the
\texttt{inla-program} which is written in \texttt{C}. This is a
convenient feature that allows us to prototype new models easily,
despite the fact that the fGn model does not have any Markov
properties. In general, Markov properties would improve the
computational efficiency significantly, see \citet[Ch.~2]{book80}.

The \texttt{rgeneric} code to represent the fGn model is given below.
We make use of the Toeplitz structure of the covariance matrix to
speed up the computations. The models is defined as a function
returning properties of the Gaussian model, like its graph, the
precision matrix, the normalising constant, and so on.
\begin{verbatim}
rgeneric.fgn = function (cmd = c("graph", "Q", "initial",
                                 "log.norm.const", "log.prior",
                                 "quit", "mu", "theta.fun"),
                         theta = NULL,
                         args = NULL) 
{
    stopifnot(require(ltsa) && require(FGN) && require(HKprocess))
    
    theta.fun = function(...) {
        return(list(prec = function (x) exp(x),
                       H = function(x) exp(x)/(1 + exp(x))))
    }
    interpret.theta = function(n, theta) {
        fun = theta.fun()
        return(list(prec = fun[[1]](theta[1]),
                       H = fun[[2]](theta[2])))
    }
    graph = function(n, theta) {
        return (inla.as.sparse(matrix(1, n, n)))
    }
    Q = function(n, theta) {
        param = interpret.theta(n, theta)
        S = toeplitz(acvfFGN(param$H, n-1))
        S = inla.as.sparse(param$prec * TrenchInverse(S))
        return(S)
    }
    log.norm.const = function(n, theta) {
        param = interpret.theta(n, theta)
        r = acvfFGN(param$H, n-1) 
        S = toeplitz(r) / param$prec
        val = ltzc(r, rep(0, n))
        val2 = c(-n/2 * log(2*pi) -0.5*val[2] + n/2 * log(param$prec))
        return(val2)
    }
    log.prior = function(n, theta) {
        param = interpret.theta(n, theta)
        lprior = inla.pc.dprec(param$prec, u=1, alpha=0.01, log=TRUE)
                 + log(param$prec)
        lprior = lprior + PCPRIOR.H(theta[2])
        return (lprior)
    }
    mu = function(n, theta) return (numeric(0))
    initial = function(n, theta) return(c(4, 0.))
    quit = function(n, theta) return(invisible())

    ## the PCPRIOR.H-function must be provided
    stopifnot(exists("PCPRIOR.H") && is.function(PCPRIOR.H))
    cmd = match.arg(cmd)
    val = do.call(cmd, args = list(n = as.integer(args$n),
                      theta = theta))

    return(val)
}
\end{verbatim}

We can now include the given fGn model component together with other
latent components. This is an easy example, where the function
returning the log-prior for $\rho$ (the internal representation of
$H$) is defined in the file \texttt{fgn-prior.R} (not included here).
\begin{verbatim}
library(FGN)
n=200; H=0.8
source("fgn.R")
model = inla.rgeneric.define(rgeneric.fgn, n=n,
                             R.init = "fgn-prior.R")
y = scale(SimulateFGN(n,H))
r = inla(y ~ -1 + f(idx, model=model), verbose=TRUE, 
    data = data.frame(y = y, idx = 1:n, iidx = 1:n),
    control.family = list(hyper = list(
                           prec = list(initial = 12, fixed=TRUE))))
summary(r)
\end{verbatim}

\end{document}